\documentclass[twocolumn,nofootinbib,showpacs,preprintnumbers,amsmath,amssymb,superscriptaddress,aps,pra,10pt]{revtex4-1}

\usepackage{amsmath}
\usepackage{amssymb}
\usepackage{amsfonts}
\usepackage{amsthm}
\usepackage{graphicx}
\usepackage[pdftitle={Non-stoquastic-quantum}]{hyperref}
\usepackage{bm}

\usepackage{newtxtext,newtxmath}

\makeatletter

\newcommand{\ket}[1]{|{#1}\rangle}

\newcommand{\bra}[1]{\langle{#1}|}
\newcommand{\bkt}[2]{\langle{#1}|{#2}\rangle}

\newcommand{\TF}{{\rm TF}}
\newcommand{\SC}{{\rm SC}}

\newcommand{\GS}{{\rm GS}}
\renewcommand{\H}{\hat{H}}
\newcommand{\V}{\hat{V}}
\newcommand{\s}{\hat{\sigma}}
\renewcommand{\S}{\hat{S}}

\newcommand{\sbigotimes}{%
  \mathop{\mathchoice{\textstyle\bigotimes}{\bigotimes}{\bigotimes}{\bigotimes}}%
}

\begin{document}
\title{Relation between quantum fluctuations and the performance enhancement of\\ quantum annealing in a nonstoquastic Hamiltonian}
\author{Yuki Susa}
\email{susa@stat.phys.titech.ac.jp}
\affiliation{Department of Physics, Tokyo Institute of Technology, Oh-okayama, Meguro-ku, Tokyo, 152-8551, Japan}
\author{Johann F. Jadebeck}
\email{fredrik.jadebeck@rwth-aachen.de}
\affiliation{Department of Physics, Tokyo Institute of Technology, Oh-okayama, Meguro-ku, Tokyo, 152-8551, Japan}
\affiliation{Department of Physics, RWTH Aachen University, Templergraben 55, 52056 Aachen, Germany}
\author{Hidetoshi Nishimori}
\email{nishimori@phys.titech.ac.jp}
\affiliation{Department of Physics, Tokyo Institute of Technology, Oh-okayama, Meguro-ku, Tokyo, 152-8551, Japan}
\date{\today}
\begin{abstract}
We study the relation between quantum fluctuations and the significant enhancement of the performance of quantum annealing in a mean-field Hamiltonian.  First-order quantum phase transitions were shown to be reduced to second order by antiferromagnetic transverse interactions in a mean-field-type many-body-interacting Ising spin system in a transverse field, which means an exponential speedup of quantum annealing by adiabatic quantum computation. We investigate if and how quantum effects manifest themselves around these first- and second-order phase transitions to understand if the antiferromagnetic transverse interactions appended to the conventional transverse-field Ising model induce notable quantum effects. By measuring the proximity of the semiclassical spin-coherent state to the true ground state as well as the magnitude of the concurrence representing entanglement, we conclude that significant quantum fluctuations exist around second-order transitions, whereas quantum effects are much less prominent at first-order transitions. Although the location of the transition point can be predicted by the classical picture, system properties near the transition need quantum-mechanical descriptions for a second-order transition but not necessarily for first order.  It is also found that quantum fluctuations are large within the ferromagnetic phase after a second-order transition from the paramagnetic phase. These results suggest that the antiferromagnetic transverse interactions induce marked quantum effects, and this fact would be related to closely to the significant enhancement of the performance of quantum annealing.
\end{abstract}
\maketitle

\section{Introduction}

%
Combinatorial optimization problems~\cite{Garey1979} are generally hard to solve since the computational complexity (the time necessary to reach the solution) is typically exponential in the problem size~\cite{Barahona1982}.  Quantum annealing (QA)~\cite{Kadowaki1998,Kadowaki1998b,Finnila1994,Brooke1999,Santoro2002,Santoro2006,Das2008,Morita2008} has been devised as a metaheuristic for combinatorial optimization problems and uses quantum fluctuations for state search in place of thermal fluctuations for the classical counterpart of simulated annealing~\cite{Kirkpatrick1983}. In its formulation as quantum adiabatic computing~\cite{Farhi2001}, the system is supposed to follow the instantaneous ground state of a quantum system, typically the transverse-field Ising model, and the final state is expected to be close to the ground state of a classical Ising model which encodes the combinatorial optimization problem that one wishes to solve.

Since the system is expected to follow the instantaneous ground state as faithfully as possible, it is desirable that the energy gap between the ground state and the first excited state is not very small, as a small gap causes a transition between states. In particular, if the system encounters a quantum phase transition in the course of annealing, it causes a problem because the gap vanishes in the thermodynamic limit. Therefore one should carefully check the existence and the type of a possible phase transition. It is generally the case that a first-order quantum phase transition is characterized by a gap closing exponentially as a function of the system size. This causes a difficulty for QA because the time necessary to follow the instantaneous ground state depends polynomially on the inverse of the energy gap according to the adiabatic theorem of quantum mechanics~\cite{Albash2016}. In contrast, at a second-order transition, the gap closes polynomially, and thus the computation time grows polynomially as the system size increases.  This is a situation where the problem is considered to be solved easily.

It was pointed out in Ref.~\cite{Jorg2010} that the simple ferromagnetic Ising model with $p$-body infinite-range interactions has a first-order phase transition as a function of the strength of the transverse field.  This implies that the trivial ferromagnetic ground state cannot be reached easily by QA.  This serious problem has been shown to be circumvented in Refs.~\cite{Seki2012,Seoane2012} by the introduction of antiferromagnetic transverse interactions into the Hamiltonian, by which  first-order phase transitions are reduced to second order. A similar phenomenon has been observed in the Hopfield model \cite{Seki2015}.

Closely related to the above-mentioned phenomenon of the change of transition order is the concept of a stoquastic or nonstoquastic Hamiltonian. A stoquastic Hamiltonian is defined as a Hamiltonian operator whose matrix representation has its off-diagonal elements all real and nonpositive in an appropriate basis, usually the computational basis to diagonalize the $z$ component of the Pauli operator $\hat{\sigma}_i^z$ at each site $i$~\cite{Bravyi2008}. This means that a system with a stoquastic Hamiltonian can be simulated classically by using the Suzuki-Trotter decomposition~\cite{Suzuki1976}.\footnote{See, however, Ref.~\cite{Albash2016} for exceptions.}  A nonstoquastic Hamiltonian violates the above condition of real and nonpositive off-diagonal matrix elements. It is often difficult to classically simulate a system with a nonstoquastic Hamiltonian because of the sign problem in the effective Boltzmann factor~\cite{Albash2016}.  A Hamiltonian can be stoquastic or nonstoquastic depending on the choice of the basis of representation~\cite{Bravyi2008}. In the context of quantum annealing, it is natural to focus our attention on the representation to diagonalize $\hat{\sigma}_i^z$ because the state of the final Hamiltonian, the classical Ising model, is measured in this basis. We use this convention in this paper. In this sense, the transverse-field Ising model with ferromagnetic $p$-body infinite-range interactions is nonstoquastic if antiferromagnetic transverse interactions are introduced.\footnote{It is useful to remember that the Hopfield model in a  transverse field also becomes nonstoquastic in the $\hat{\sigma}_i^z$ basis by the introduction of antiferromagnetic transverse interactions \cite{Seki2015}, where the antiferromagnetic transverse interactions play a role similar to that in the simple ferromagnetic model to reduce first-order transitions to second order under certain conditions. The Hopfield model with antiferromagnetic transverse interactions cannot be made stoquastic by an exchange of $x$ and $z$ axes, in contrast to the ferromagnetic $p$-spin model.}

The reduction of first-order transitions to second order by antiferromagnetic transverse interactions in the $p$-body interacting ferromagnetic Ising model suggests that a nonstoquastic Hamiltonian that cannot be simulated easily classically would potentially have enhanced efficiency of quantum annealing significantly, which is obviously a remarkable effect.  See also Ref.~\cite{Nishimori2016} for related discussions.  Numerical evidence has been provided that nonstoquastic Hamiltonians may enhance the efficiency of quantum annealing~\cite{Farhi2002_2,Crosson2014,Hormozi2016}.

The results in the analytical studies in Refs.~\cite{Seki2012,Seoane2012} apply mostly to the behavior of the system in the thermodynamic limit, where the spin operators appearing in the Hamiltonian are almost classical because they are written as the sums of all microscopic spin variables and therefore become macroscopic in size. Although the reduction of first-order transitions to second order by antiferromagnetic transverse interactions can be correctly described by statistical-mechanical calculations valid in the thermodynamic limit, where the (semi)classical picture is justified, it is not easy to extract subtle quantum effects, which may (or may not) play a crucial role, from direct statistical-mechanical computations. It is therefore interesting and important to understand in more detail if and how quantum effects affect the properties of the mean-field model with a nonstoquastic Hamiltonian.  In other words, we are interested in the subtle interplay between the dominantly classical characteristics of the mean-field model and the quantum fluctuations caused by antiferromagnetic transverse interactions in large but finite-size systems.

In the present paper we study this problem in the transverse-field Ising model with infinite-range $p$-spin ferromagnetic interactions ($p$-spin model). We follow Ref.~\cite{Muthukrishnan2016} to study the trace-norm distance between the true ground state and the semiclassical spin-coherent state for large but finite-size systems. Similarities and differences between semiclassical and true ground states are clearly understood through this analysis. We also compute the concurrence, representing the degree of entanglement between two spins in the system. It is shown that the concurrence behaves differently depending on the order of transition. These results lend support to the proposition that the reduction of the order of transition, caused by antiferromagnetic transverse interactions, takes place simultaneously with the emergence of conspicuous quantum behavior of the system. This observation indicates a way to understand how quantum effects would enhance the efficiency of quantum annealing through a nonstoquastic Hamiltonian.

This paper is structured as follows. In Sec. \ref{sec:p-spin_model}, we formulate the Hamiltonian for QA in the $p$-spin model with antiferromagnetic transverse interactions. Section \ref{sec:trace-norm_dist} shows the behavior of the trace-norm distance, and Sec. \ref{sec:concurrence} is devoted to the analyses of concurrence.  Discussion is given in Sec. \ref{sec:summary}.  Details of some of the calculations are described in the Appendix. 

\section{Ferromagnetic $p$-spin model with antiferromagnetic transverse interactions}
\label{sec:p-spin_model}

We recapitulate the general form of QA and the model that we consider in this study. The Hamiltonian for QA is written as
\begin{align}
\H(s)=(1-s)\H_i+s\H_0,
\end{align}
where $\H_{i}$ and $\H_{0}$ are the initial and final target Hamiltonians, respectively. The parameter $s$ changes from the initial value of $s=0$ to the final $s=1$. A conventional choice of the initial Hamiltonian is the transverse field,
\begin{align}
\H_i=\V_{\TF}=-\sum_{i=1}^{N}\s_{i}^{x},
\end{align}
where $\s_{i}^{x}=|0\rangle_i\langle1|+|1\rangle_i\langle0|$ is the $x$ component of the Pauli operator and the site index $i$ runs from 1 to $N$. The ground state of $\V_{\TF}$ is the trivial product state $\otimes_{i=1}^{N}(\ket{0}_i+\ket{1}_i)/\sqrt{2}$.

Our final Hamiltonian is the ferromagnetic $p$-spin model,
\begin{align}
\label{eq:pspin}
\H_0=-N\left(\frac{1}{N}\sum_{i=1}^N\s_i^z\right)^p,
\end{align}
where $\s_{i}^{z}=|0\rangle_i\langle0|-|1\rangle_i\langle1|$ is the $z$ component of the Pauli operator. For odd $p$, the ground state of $\H_0$ is $\otimes_{i=1}^{N}\ket{0}_i$, and for even $p$, the ground state is doubly degenerate, $\otimes_{i=1}^{N}\ket{0}_i$  and $\otimes_{i=1}^{N}\ket{1}_i$. This $p$-spin model reduces to the Grover problem in the limit of infinite $p$~\cite{Jorg2010}. We assume that $p$ is odd in the present paper to avoid possible inessential complications coming from the trivial degeneracy.

As noted in Ref. \cite{Nishimori2016}, the Hamiltonian (\ref{eq:pspin}) is a simple polynomial of the order parameter, and the ground state can be easily found by the method of gradient descent.  Nevertheless, the problem becomes nontrivial if we apply simulated or quantum annealing, the latter with a simple transverse field, since an effective energy barrier separates two coexisting states, ferromagnetic and paramagnetic, at the transition point and thus the phase transition is of first order. Our focus in the present paper is to investigate how the reduction of this energy barrier for quantum annealing is related to quantum effects.

As proposed in Refs.~\cite{Seki2012,Seoane2012}, we introduce the following antiferromagnetic transverse interactions into the Hamiltonian:
\begin{align}
\V_{\rm AFI}=N\left(\frac{1}{N}\sum_{i=1}^N \s_i^x\right)^2.
\end{align}
The total Hamiltonian $\H(s)$ is written as
\begin{align}
\label{eq:total_hamiltonian}
\H(s,\lambda)=s\left(\lambda \H_0+(1-\lambda)\V_{\rm AFI} \right)+(1-s)\V_{\TF},
\end{align}
using an extra parameter $\lambda$. This $\lambda$ can have an arbitrary value initially ($s=0$) because $\lambda$ disappears from the Hamiltonian when $s=0$, and it finally reaches $\lambda=1$ as $s$ approaches 1. 

Since the total spin is conserved in this Hamiltonian and the initial ground state belongs to the subspace with the largest value of the total spin, we restrict our analysis to this subspace. This fact enormously facilitates numerical computations since the dimension of the relevant Hilbert space is linear in $N$, in contrast to $2^N$ of the full Hilbert space. In the Appendix, we show the explicit matrix representation of the Hamiltonian (\ref{eq:total_hamiltonian}) in this subspace.

\section{Semiclassical analysis and the trace-norm distance}
\label{sec:trace-norm_dist}
Closely following Ref.~\cite{Muthukrishnan2016}, we measure the trace-norm distance between the exact ground state and the semiclassical spin-coherent state for large but finite-size systems to see how far the semiclassical description is accurate. 

\subsection{Semiclassical analysis by the spin-coherent state}

We first discuss the phase diagram using the semiclassical spin-coherent state defined as the product state,
\begin{align}
\label{eq:spin_coherent_state}
\ket{\theta,\phi}=\sbigotimes_{i=1}^N \left[\cos\left(\frac{\theta}{2}\right)\ket{0}_i+\sin\left(\frac{\theta}{2}\right)e^{i\phi}\ket{1}_i\right],
\end{align}
where all spins are assumed to have the same angular variables $\theta$ and $\phi$. 

The semiclassical potential is defined as the expectation value of the Hamiltonian by the spin-coherent state divided by the system size \cite{Muthukrishnan2016,Farhi2002_3,Schaller2010,Boixo2016},
\begin{align}
V_{\SC}(\theta,\phi,s,\lambda)
&=\lim_{N\rightarrow\infty}N^{-1}\bra{\theta,\phi}\hat{H}(s,\lambda)\ket{\theta,\phi} \notag \\
&=s[-\lambda \cos^p\theta+(1-\lambda)\sin^2\theta \cos^2\phi] \notag \\
&~~~-(1-s)\sin\theta\cos\phi.
\end{align}
This semiclassical potential is also obtained from the Hamiltonian (\ref{eq:total_hamiltonian}) when we replace the spin operator $\sum_i \hat{\sigma}_i^z/N$ by the polar coordinate representation of a classical unit vector, $\cos\theta$, and $\sum_i \hat{\sigma}_i^x/N$ by $\sin\theta\cos\phi$. It is clearly seen that the ground state has $\phi =0$.

\begin{figure}[t]
  \centering
  \includegraphics[width=8.6cm]{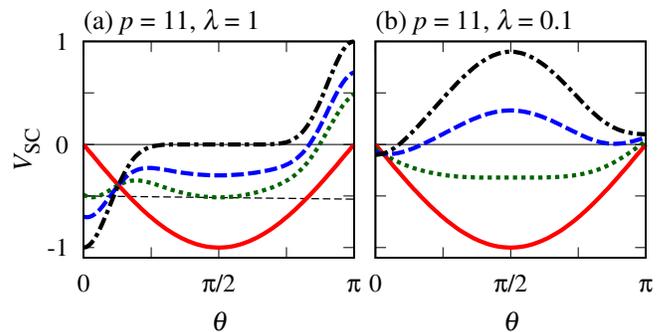}
 \caption{The semiclassical potential $V_{\SC}(\theta,\phi,s,\lambda)$ as a function of $\theta$ for $p=11$ with $\phi=0$. We fix $\lambda=1$ in (a) for the stoquastic Hamiltonian and $\lambda=0.1$ in (b) for a nonstoquastic case. In both panels, the red solid curves are for $s=0$, blue dashed curves are for $s=0.7$, and black curves are for $s=1$. The green dotted curves are for $s=0.487$ in (a) and $s=0.357$ in (b), when the first- and second-order phase transitions take place, respectively. The horizontal dashed line in (a) indicates the degenerate ground-state energy at $s=0.487$.
 }
\label{fig:semi-classical potential}
\end{figure}
Figure \ref{fig:semi-classical potential} shows the semiclassical potential  $V_{\SC}$ as a function of $\theta$ for $p=11$ with $\phi=0$. The angle $\theta$ to minimize the semiclassical potential will be denoted as $\theta_{\min}$. In both Figs. \ref{fig:semi-classical potential}(a) and \ref{fig:semi-classical potential}(b), $\theta_{\min}=\pi/2$ at $s=0$ and $\theta_{\min}=0$ at $s=1$.

Figure \ref{fig:semi-classical potential}(a) is for the stoquastic Hamiltonian with $\lambda=1$. Here, there exists a first-order transition at $s=0.487$, where $\theta_{\min}$ shows a jump. Figure \ref{fig:semi-classical potential}(b) has $\lambda =0.1$ for a nonstoquastic case.  The transition at $s=0.357$ is of second order, where $\theta_{\min}$ starts to change continuously away from $\pi/2$.

\begin{figure}[b]
  \centering
  \includegraphics[width=8.6cm]{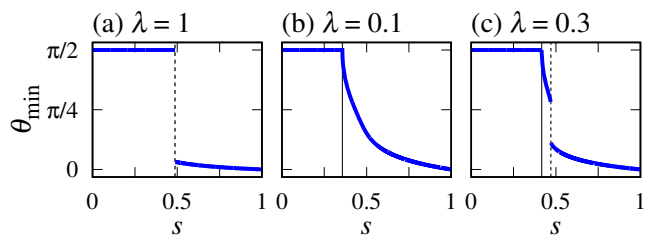}
 \caption{The optimal angle $\theta_{\min}$ as a function of $s$ for (a) $\lambda=1$, (b) $\lambda=0.1$, and (c) $\lambda=0.3$ when $p=11$ and $\phi=0$. The vertical dotted and solid lines show the first- and second-order phase transitions, respectively. First-order phase transitions take place at $s=0.487$ in (a) and $s=0.471$ in (c).  A second-order transition exists at $s=0.357$ in (b) and at $s=0.417$ in (c).}
 \label{fig:theta_min}
\end{figure}

\begin{figure*}[t]
  \centering
  \includegraphics[width=14cm]{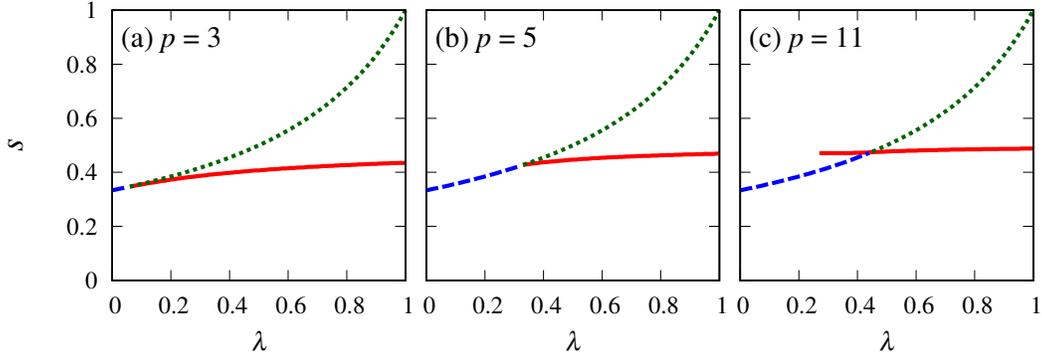}
 \caption{Phase diagrams for three values of $p$: (a) $p=3$, (b) $p=5$, and (c) $p=11$. The red solid and blue dashed curves represent first- and second-order phase transitions, respectively. The blue dashed curve satisfies $s=1/(3-2\lambda)$, so does the green dotted curve, the latter of which is, however, not a line of phase transitions.}
 \label{fig:phase_diagram}
\end{figure*}
\begin{figure*}[t]
  \centering
  \includegraphics[width=\textwidth]{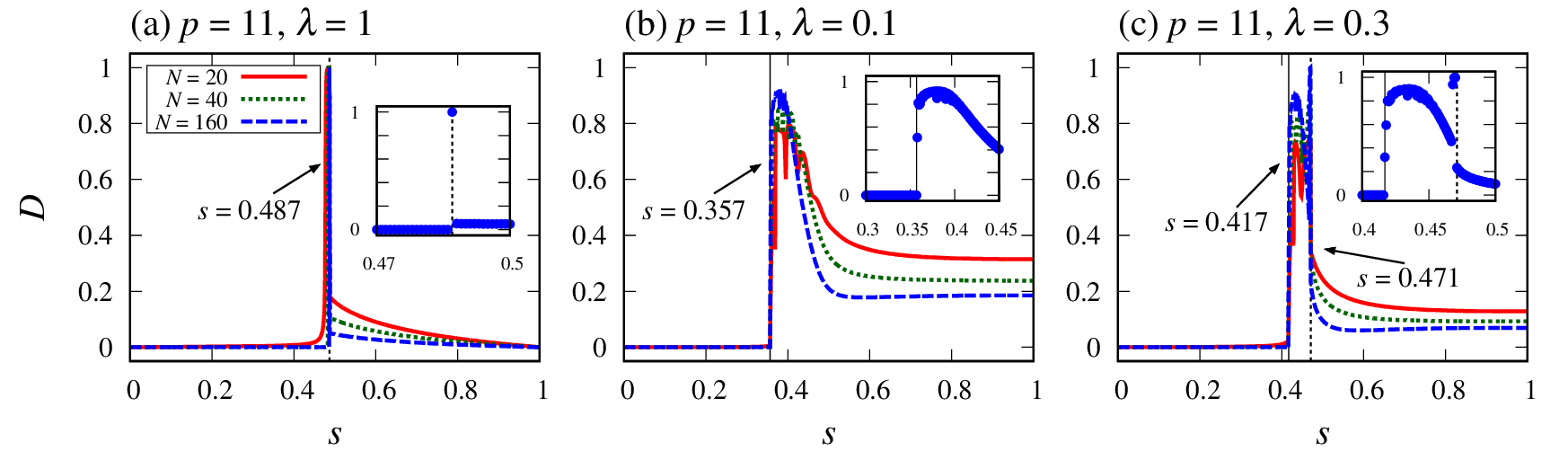}
 \includegraphics[width=\textwidth]{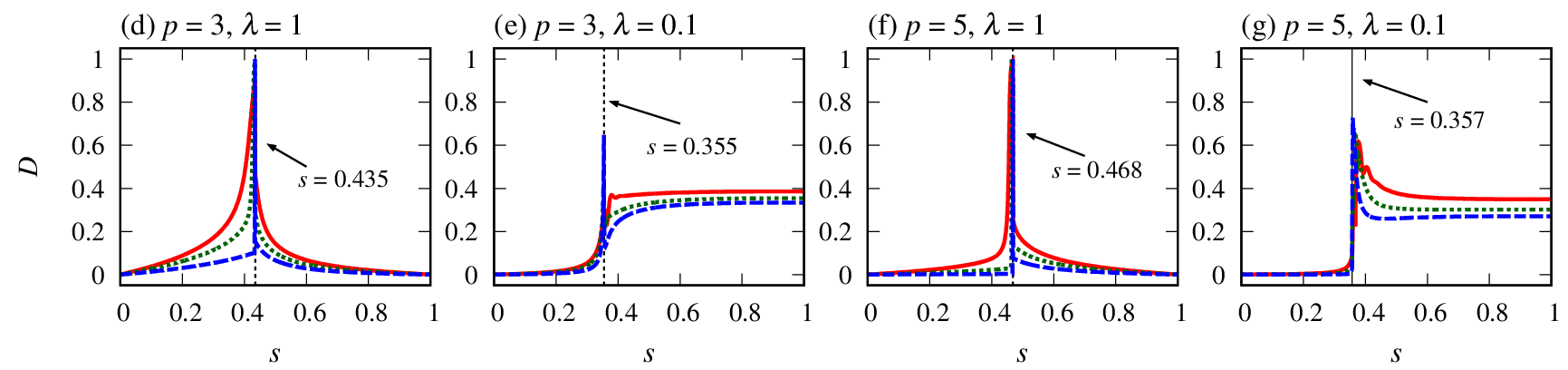}
 \caption{Trace-norm distance $D$ between the spin-coherent state $\ket{\theta_{\min},0}$ and the exact ground state $\ket{\psi_{\GS}}$ as a function of $s$. Different system sizes are coded by colors, $N=20$ (red solid), $N=40$ (green dotted), and $N=160$ (blue dashed). Insets  are for the largest size $N=160$ around the phase transition. The vertical dotted and solid lines are the first- and second-order phase transitions, respectively.}
 \label{fig:trace_norm_dist}
\end{figure*}
Figure \ref{fig:theta_min} shows $\theta_{\min}$ for $p=11$ for three values of $\lambda$. In Fig. \ref{fig:theta_min}(a) for $\lambda=1$, $\theta_{\min}$ jumps at $s=0.487$, indicating a first-order transition, whereas in Fig. \ref{fig:theta_min}(b), $\theta_{\min}$ starts to change continuously at $s=0.357$, a signature of a second-order transition. Figure \ref{fig:theta_min}(c) for $\lambda =0.3$ shows the interesting case with both first- and second-order transitions appearing at $s=0.471$ and $s=0.417$, respectively.

Figure \ref{fig:phase_diagram} displays the phase diagrams obtained from the semiclassical potential. The red and blue curves are for first- and second-order phase transitions, respectively. The latter can be evaluated by the condition
\begin{align}
\left.\frac{\partial^2 V_{\SC}(\theta,\phi,s,\lambda) }{\partial \theta^2}\right|_{\theta=\pi/2,\phi=0}&=-2s(1-\lambda)+(1-s)=0,
\end{align}
which gives $s=1/(3-2\lambda)$.  The green dotted curve shows this equation after the second-order transition drawn in blue is replaced by a first-order line in red and therefore does not represent a line of phase transitions.

As seen in Figs. \ref{fig:phase_diagram}(b) and \ref{fig:phase_diagram}(c) for $p=5$ and 11, respectively, there exists a path from the initial state at $s=0$ ($\lambda$ arbitrary) to the final $s=\lambda =1$ that avoids first-order transitions (red curves). In contrast, when $p=3$, it is impossible to avoid first-order phase transitions.  These results agree with those by full quantum statistical-mechanical computations for both first- and second-order transition lines \cite{Seki2012}.

\subsection{Trace-norm distance}

Next, we study the trace-norm distance between the spin-coherent state $\ket{\theta_{\min},0}$ minimizing the semiclassical potential and the exact ground state $\ket{\psi_{\GS}}$ of the Hamiltonian (\ref{eq:total_hamiltonian}) for finite-size systems to see if and when the semiclassical spin-coherent state is close to the true ground state.  The trace-norm distance is defined by
\begin{align}
 \label{eq:trace_norm_dist}
D = \sqrt{1-|\bkt{\theta_{\min},0}{\psi_{\GS}}|^2}.
\end{align}
The trace-norm distance is $D=0$ when $\ket{\theta_{\min},0}$ coincides with $\ket{\psi_{\GS}}$ and is $D=1$ when they are orthogonal. Some technical details to calculate $D$ are described in the Appendix.

\begin{figure*}[t]
  \centering
  \includegraphics[width=\textwidth]{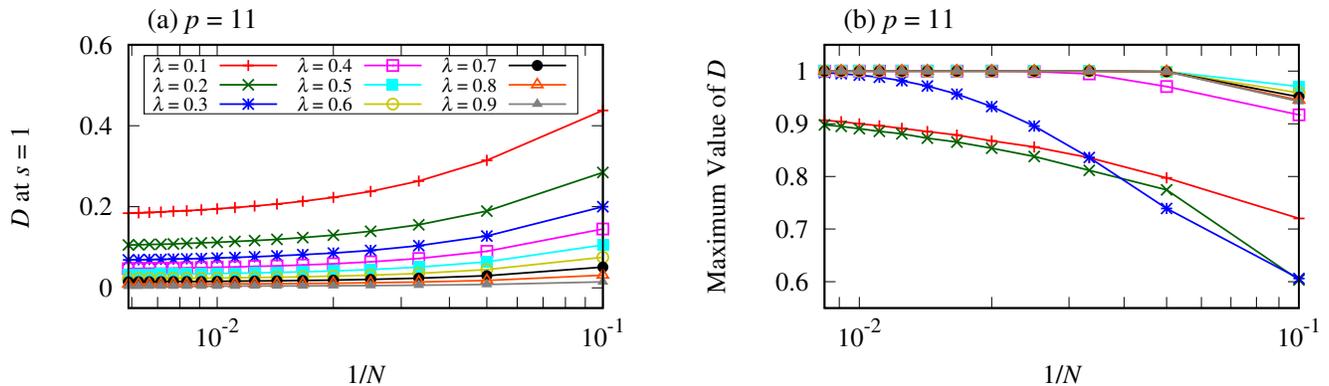}
 \caption{(a) System-size dependence of the trace-norm distance at $s=1$ and (b) the value at its maximum for $p=11$.  Curves in (b) are distinguished by the same color code as in (a).}
 \label{fig:trace_norm_dist_2}
\end{figure*}

Figure \ref{fig:trace_norm_dist} displays the trace-norm distance for $p=3, 5$, and 11.  The distance $D$ is sharply peaked around the transition points. The height of a peak is smaller at a second-order phase transition [vertical solid lines in Figs. \ref{fig:trace_norm_dist}(b), \ref{fig:trace_norm_dist}(c), and \ref{fig:trace_norm_dist}(g)] than at a first-order phase transition [vertical dotted lines in Figs. \ref{fig:trace_norm_dist}(a), \ref{fig:trace_norm_dist}(c), \ref{fig:trace_norm_dist}(d), \ref{fig:trace_norm_dist}(e), and \ref{fig:trace_norm_dist}(f)]. The latter has $D\approx1$, which means that the true ground state is completely different from the semiclassical state.  The value at the peak is plotted as a function of the inverse system size in Fig. \ref{fig:trace_norm_dist_2}(b). The peak value clearly approaches 1 for first-order transitions ($\lambda \ge 0.3$), whereas it is slower to increase at second-order transitions ($\lambda =0.1$ and 0.2).

In Fig. \ref{fig:trace_norm_dist}, the peak is very sharp at a first-order transition and is relatively broad for second order. The distance $D$ remains finite in the ferromagnetic phase (large $s$) for the smaller value of $\lambda =0.1$ [Figs. \ref{fig:trace_norm_dist}(b), \ref{fig:trace_norm_dist}(e), and \ref{fig:trace_norm_dist}(g)] but drops rapidly to zero for $\lambda =1$ [Figs. \ref{fig:trace_norm_dist}(a), \ref{fig:trace_norm_dist}(d), and \ref{fig:trace_norm_dist}(f)].  The size dependence of the trace-norm distance at $s=1$ is depicted in Fig. \ref{fig:trace_norm_dist_2}(a).  Here, we see quantitative differences among different values of $\lambda$, but it is not necessarily clear whether or not there exists a qualitative difference. It is at least true that the ferromagnetic phase for small $\lambda$ (i.e.,  a large coefficient of antiferromagnetic transverse interactions) has a ground state significantly different from the spin-coherent state.

These results show that the state of the system is far from the spin-coherent state around the transition, and such a range of the parameter $s$ is wider for second-order transitions.  The paramagnetic phase, which exists at small $s$ for any $\lambda$, is very well described by the semiclassical spin-coherent state for any $\lambda$.  However, the ferromagnetic phase, which occupies the upper half of the phase diagrams in Fig. \ref{fig:phase_diagram}, shows deviations from the spin-coherent state for small values of $\lambda$. In other words, antiferromagnetic transverse interactions $\hat{V}_{\rm AFI}$ with a large coefficient cause noticeable departures of the ground-state wave function from the semiclassical spin-coherent state at a phase transition and within the ferromagnetic phase.

It is premature to conclude that these deviations of the ground state from the spin-coherent state caused by antiferromagnetic transverse interactions are the {\em sole} origin of the reduction of a first-order transition to second order because the case of $p=3$ has no second-order transition but the ferromagnetic phase for $\lambda=0.1$ has a different state from the spin-coherent state, as is evident in Fig. \ref{fig:trace_norm_dist}(e). It is useful to remember here that the system with $p=3$ is special in the sense that a simple Landau-type argument precludes a second-order transition by the introduction of a cubic term in the Landau expansion of the free energy,
\begin{equation}
F(m)\sim a\,m^2+b\, m^3+c\, m^4+\cdots.
\end{equation}
For larger values of $p (\ge 5)$, the leading two terms of the Landau expansion would be quadratic and quartic in $m$, 
\begin{equation}
F(m)\sim a\, m^2+b\, m^4+c\, m^5+\cdots,
\end{equation}
which leads to the possibility of a second-order transition when the coefficient $c$ is small.  We may conclude that antiferromagnetic transverse interactions, which make the total Hamiltonian nonstoquastic, promote deviations from the semiclassical spin-coherent state at the transition as well as in the ferromagnetic phase, and the order of transition may be reduced from first to second when a few additional conditions are satisfied (or a few detrimental conditions are removed).

\section{Entanglement and energy spectrum}
\label{sec:concurrence}
\begin{figure*}[tb]
  \centering
  \includegraphics[width=\textwidth]{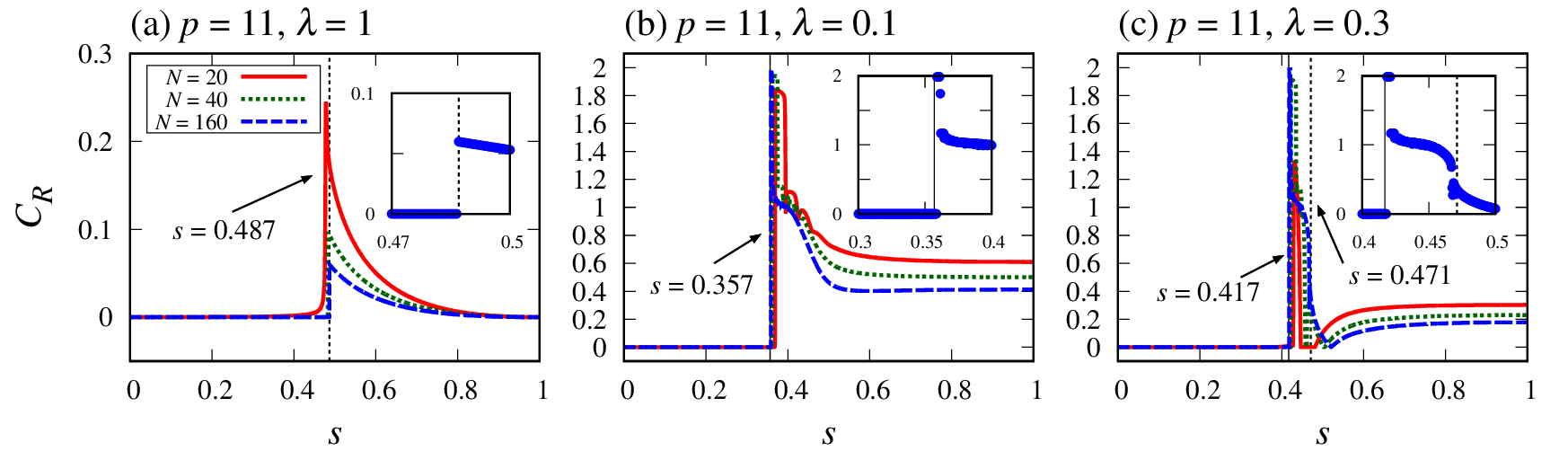}
  \includegraphics[width=\textwidth]{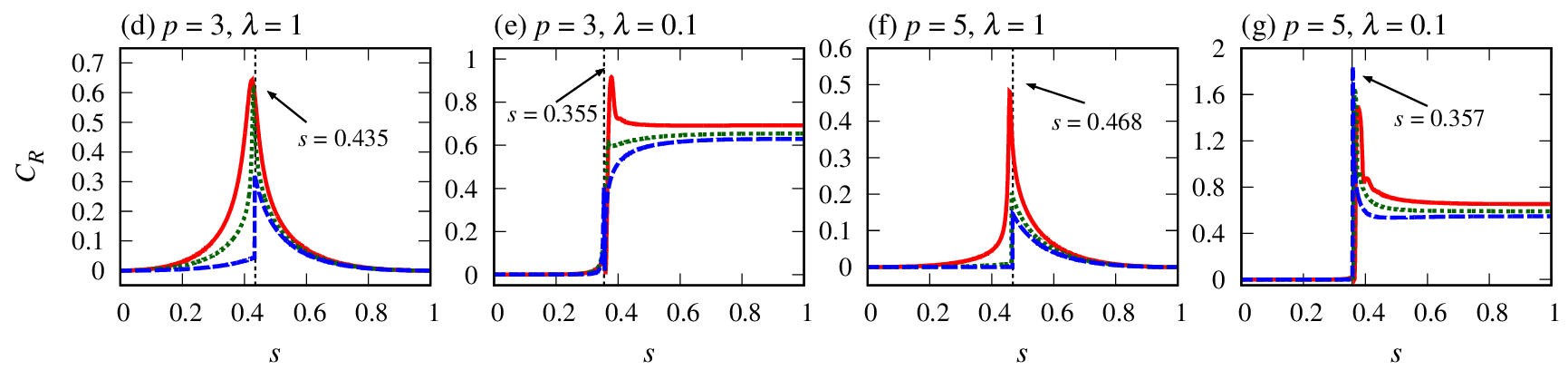}
 \caption{Rescaled concurrence $C_R$ as a function of $s$ for fixed $\lambda$ for (a)--(c) $p=11$, (d) and (e) $p=3$, and (f) and (g) $p=5$. The system size is color-coded as $N=20$ (red solid curve), $N=40$ (green dotted curve), and $N=160$ (blue dashed curve). Insets are for $N=160$ around the phase transition point. The vertical dotted and solid lines are the first- and second-order phase transitions, respectively.}
 \label{fig:concurrence_1}
\end{figure*}

To further understand the role of quantum effects in the nonstoquastic Hamiltonian, we evaluate the degree of entanglement represented by the concurrence.  This is a quantity to measure how much two sites are entangled in a system. The stoquastic case has already been discussed in detail in Ref. \cite{Filippone2011}, where it was found that three different measures of entanglement, i.e., concurrence, entanglement entropy, and negativity, behave similarly in the sense that, when the concurrence has a jump at a first-order transition, so do the other two measures.  Also, a sharp peak in the concurrence is observed when the entanglement entropy is divergent or the negativity is sharply peaked at a second-order transition. Thus the type of transition and the degree of quantum fluctuations can be inferred by any one of the three measures in the present mean-field-type model with infinite-range interactions \cite{comment_Filippone}. This is a deviation from the case of finite dimensions, where the entanglement entropy has a distinct feature to detect if the entanglement extends beyond the interface region between two parts of the system.  We use the concurrence, not the more standard entanglement entropy, to distinguish different types of behavior of entanglement because the concurrence, which remains finite at a second-order transition, has weaker dependence on the system parameters (size and coefficients in the Hamiltonian), and it is therefore harder to detect qualitative differences for different types of phase transitions by the concurrence.  If we were able to see clear qualitative differences in the behavior of the concurrence between different values of $\lambda$, that would serve as clear evidence that the entanglement plays different roles in different types of transitions.

\begin{figure*}[tb]
  \centering
  \includegraphics[width=\textwidth]{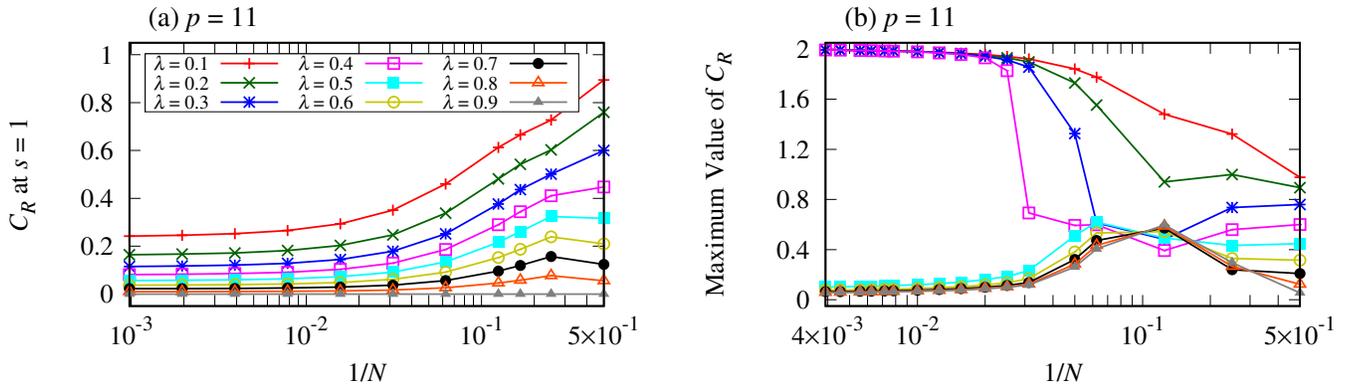}
 \caption{(a) Rescaled concurrence $C_R$ at $s=1$ as a function of $1/N$ for $p=11$. (b) The maximum value of $C_R$ for $s$ as a function of $1/N$. Both plots are in a semilog scale. The same color code for $\lambda$ is used in both panels.}
 \label{fig:concurrence_2}
\end{figure*}

The concurrence $C(\rho)$, for a given density matrix $\rho$, is defined by~\cite{Hill1997,Wootters1998, Yano2005}
\begin{align}
C(\rho)=\max \{0,\lambda_1-\lambda_2-\lambda_3-\lambda_4 \},
\end{align}
where $\lambda_{1}, \lambda_{2}, \lambda_{3}$, and $\lambda_{4}$ are the square roots of the eigenvalues, in decreasing order, of the product matrix $R=\rho(\s_i^y\otimes\s_j^y)\rho^{\ast}(\s_i^y\otimes\s_j^y)$, with $\rho^{\ast}$ being the complex conjugate of $\rho$.  Since our Hamiltonian (\ref{eq:total_hamiltonian}) has infinite-range interactions, all sites are equivalent, and thus we may choose any pair of sites $(i, j)$ to calculate the concurrence. Following Refs.~\cite{Wang2003,Vidal2004,Vidal2006,Filippone2011}, we consider the rescaled concurrence $C_R(\rho)$ defined by
\begin{align}
C_R(\rho)=(N-1)C(\rho) \label{rescaled_concurrence}
\end{align}
to extract nontrivial values. Since the system is likely to behave classically in the leading order of system size, i.e., in the thermodynamic limit, the value of concurrence vanishes in the leading order.  We should evaluate the next-order correction to identify the subtle quantum symptoms, which is reflected in the definition of the rescaled concurrence in Eq. (\ref{rescaled_concurrence}).

Figure \ref{fig:concurrence_1} shows the rescaled concurrence as a function of $s$. The behavior may look similar to the corresponding data for the trace-norm distance in Fig. \ref{fig:trace_norm_dist}, but there exist subtle but important differences.  It is seen in Figs. \ref{fig:concurrence_1}(b), \ref{fig:concurrence_1}(c), \ref{fig:concurrence_1}(e), and \ref{fig:concurrence_1}(g) that the rescaled concurrence assumes nonzero values in the ferromagnetic phase for the nonstoquastic case.  To check if those finite values persist in the limit of large system size at $s=1$, we plot in Fig. \ref{fig:concurrence_2}(a) the rescaled concurrence as a function of the inverse of the system size for various values of $\lambda$.  They all indicate a finite value in the limit of large system size, and the limiting value is a monotonic function of $\lambda$, with a larger value for smaller $\lambda$.  Thus antiferromagnetic transverse interactions enhance entanglement not just at a transition point but also within the ferromagnetic phase for small $\lambda$ (larger coefficient of antiferromagnetic transverse interactions), as expected from the analysis of the trace-norm distance.

Figure \ref{fig:concurrence_2}(b) shows the maximum value of the rescaled concurrence observed at the transition point plotted as a function of $1/N$. It is seen that the tendency is clearly classified as two types.  The maximum increases monotonically and approaches a finite value close to 2 for second-order transitions ($\lambda \le 0.4$), whereas it decreases toward a very small value in the first-order cases ($\lambda \ge 0.5$). This is an important {\em qualitative} difference between the two cases. This definitely shows that a second-order transition is characterized by a finite amount of entanglement, whereas quantum effects are much weaker, almost negligible for $p=11$, at a first-order transition.

\begin{figure}[tb]
  \centering
  \includegraphics[width=8cm]{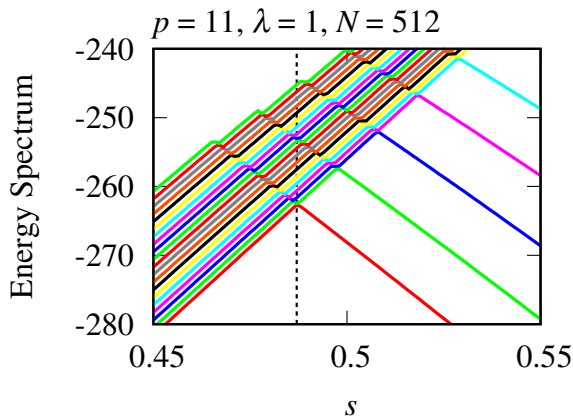}
 \caption{Instantaneous energy spectrum for $p=11$ around the first-order transition at $s=0.487$ (vertical dotted line). It is observed that a cascade of diabatic transitions would excite the system from the  ground state at $s<0.487$ to higher and higher excited states at $s>0.487$.}
 \label{fig:eigen_energy_p11_n512}
\end{figure}

It was shown in Ref. \cite{Wu2004} that a discontinuity in the concurrence at a first-order transition is closely related to a corresponding discontinuity in the derivative of the ground-state energy. Plotted in Fig. \ref{fig:eigen_energy_p11_n512} is the instantaneous energy spectrum for the case of $p=11, \lambda =1$, and $N=512$.
A clear discontinuity in the derivative is observed in the ground-state energy at the transition point at $s=0.487$, as expected from the discontinuity in the concurrence in Fig. \ref{fig:concurrence_1}(a).

Another interesting aspect of the energy spectrum is the existence of a cascade of possible diabatic transitions from the ground state at smaller $s<0.487$ toward higher and higher excited states beyond the transition point $s>0.487$.  There does not exist a backward cascade toward the ground state at a later stage of annealing (larger $s$), in contrast to the example found in Ref.~\cite{Muthukrishnan2016}.\footnote{See also Refs.~\cite{Crosson2014,Brady2016} for related observations.} In this sense, the present model in the stoquastic limit ($\lambda =1$) with a first-order transition may be taken as a hard problem to solve even by an ingenious diabatic control of the system evolution. Similar properties hold for the nonstoquastic case with a first-order transition.

\section{Discussion}
\label{sec:summary}
We have seen that antiferromagnetic transverse interactions appended to the Hamiltonian reduce first-order phase transitions to second order in the $p$-body interacting Ising model with $p\ge 5$.  The phase diagram obtained from the semiclassical potential using the spin-coherent state agrees with the phase diagram drawn by a quantum statistical-mechanical method. We can thus conclude that the location of phase transitions can be predicted by the classical analysis. This sounds natural because the Hamiltonian of Eq. (\ref{eq:total_hamiltonian}) is written only in terms of the total spin operators,
\begin{equation}
    \hat{m}_x=\frac{1}{N}\sum_{i=1}^N \hat{\sigma}_i^x,\quad
    \hat{m}_z=\frac{1}{N}\sum_{i=1}^N \hat{\sigma}_i^z,
\end{equation}
which commute in the thermodynamic limit,
\begin{equation}
\big[ \hat{m}_z, \hat{m}_x \big] =\frac{2i}{N^2}\sum_{i=1}^N \hat{\sigma}_i^y \longrightarrow 0
\quad (N\to\infty),
\end{equation}
and thus can be regarded as the corresponding components of a classical unit vector.  The reduction of a first-order transition to second order at small $\lambda$ (large coefficient of antiferromagnetic transverse interactions) can then be explained without reference to quantum mechanics. Critical fluctuations in the classical sense exist around second-order transitions.

Then, a natural question arises regarding why we need quantum mechanics in the analysis of the present model.  The answer is the finite-size effects. The annealing time of an adiabatic process crucially depends on the minimum energy gap, which is zero at a phase transition of any order in the thermodynamic limit but is finite for finite-size systems.  We therefore have to carefully scrutinize the finite-size effects around a phase transition to understand the origin of the reduction of computational complexity coming from the reduction of the order of transition from first to second.  The rescaled concurrence suits this purpose to extract subtle finite-size effects since it represents the next-to-leading-order correction in $1/N$ of the concurrence itself, as is evident in the definition of Eq. (\ref{rescaled_concurrence}).  We have seen that the maximum of the rescaled concurrence approaches a relatively large value at second-order transitions, whereas it remains very small for the first-order case.  This is a clear signature that quantum effects, entanglement, prevail around second-order transitions.  Also, the data of the trace-norm distance indicate that the semiclassical spin-coherent state is insufficient to describe the true quantum ground state around a second-order transition.  In contrast,  the trace-norm distance decays quickly away from a first-order transition point, although it is large exactly at the transition point. We thus conclude that the reduction of the order of transition shows up hand in hand with the emergence of noticeable signatures of quantum effects.

It may be appropriate to note here that the existence of significant {\em quantum} fluctuations at a transition is not necessarily a trivial consequence of the second-order transition.  Significant fluctuations already exist at a classical level, which is necessary and sufficient to characterize the classical transition as second order. Persistence of fluctuations to the subleading quantum level may be possible intuitively but should be confirmed explicitly, which we have done.

It is still under debate whether the existence of entanglement is closely related to the enhanced performance of quantum annealing with favorable \cite{Bauer2011} and unfavorable \cite{Hauke2015} examples.  We have given an interesting example using semianalytical calculations (i.e., without recourse to extensive numerical computations) in which a nonstoquastic Hamiltonian, which cannot be simulated easily classically, enhances the entanglement and, at the same time, reduces the computation time enormously from exponential (first-order transition) to polynomial (second order).  Still, further studies are necessary, as emphasized in Ref. \cite{Albash2016}.

Another interesting result is the persistence of nonclassical features, large values of the trace-norm distance and the reduced concurrence, within the ferromagnetic phase far away from the transition point when $\lambda$ is small.  We have so far been unable to explain this fact intuitively and leave it to future studies.

\vspace{0.1in}
\section*{ACKNOWLEDGMENTS}
We thank D. Lidar, T. Albash, and K. Fujii for useful comments. J.F.J. thanks RWTH Aachen for the Mirai Scholarship for support to visit Tokyo Tech. This work was funded by the ImPACT Program of the Council for Science, Technology and Innovation, Cabinet Office, Government of Japan, and by JSPS KAKENHI Grant No. 26287086.  

\appendix*

\section{MATRIX REPRESENTATION OF THE NONSTOQUASTIC HAMILTONIAN AND AN EXPANSION OF THE TRACE-NORM DISTANCE (\ref{eq:total_hamiltonian})}

The Hamiltonian (\ref{eq:total_hamiltonian}) commutes with the total spin operator $\hat{\bm{S}}^2= (\S^x)^2+(\S^y)^2+(\S^z)^2$, where $\S^{x,y,z} = \frac{1}{2}\sum_{i=1}^{N}\s_i^{x,y,z}$. Since the process of quantum annealing starts with the ground state of the transverse-field term $V_{\rm TF}$ with the largest value of ${\hat{\bm{S}}}^2=S(S+1)$, with $S=N/2$, we can restrict our computations to this subspace.

Using the standard notation of basis vectors
\begin{subequations}
\begin{align}
\hat{\bm{S}}^2\ket{S,M}&=S(S+1)\ket{S,M}, \\
\label{eq:Sz_SM}
\S^z\ket{S,M}&=M\ket{S,M}
\end{align}
\end{subequations}
and the convention
\begin{align}
\label{eq:w}
\ket{w}:=\ket{S=N/2,M=N/2-w},
\end{align}
where $w$ is and integer from zero to $N$, the matrix elements of the Hamiltonian $[H]_{w,w^{\prime}}:=\bra{w}\H(s,\lambda)\ket{w^{\prime}}$ are found to be
\begin{subequations}
\label{eq:pentadiag_hamiltonian}
\begin{align}
[H]_{w,w}=&s\left\{-\lambda N\left(1-\frac{2w}{N}\right)^p \right. \notag \\
&\left.+(1-\lambda)\left(2w-\frac{2w^2}{N}+1\right) \right\}, \\
[H]_{w,w+1}=&[H]_{w+1,w}=-(1-s)\sqrt{(N-w)(w+1)}, \\
[H]_{w,w+2}=&[H]_{w+2,w}=\frac{1}{N}s(1-\lambda) \notag \\
&\times\sqrt{(w+1)(w+2)(N-w)(N-w-1)}.
\end{align}
\end{subequations}
All other elements are zero.

To evaluate the trace-norm distance, it is convenient to expand the ground-state wave function as
\begin{align}
\ket{\psi_{\GS}}=\sum_{w=0}^{N} e_w^{\GS}\ket{w}.
\end{align} 
The state $\ket{w}$ can be decomposed to $\ket{0}$ and $\ket{1}$ as follows:
\begin{align}
\ket{w}=
\dbinom{N}{w}^{-1/2} \sum_{x:|x|=w} \ket{x},\ x=0,1.
\end{align}
The spin-coherent state corresponding to the ground state is given as $\ket{\theta_{\min},0}$ for Eq. (\ref{eq:spin_coherent_state}).
Thus we can calculate the inner product
\begin{align}
&\bkt{\theta_{\min},0}{\psi_{\GS}} \notag \\
&~=\sum_{w=0}^{N}e_{w}^{\GS}\dbinom{N}{w}^{1/2}\cos^{N-w}\left(\frac{\theta_{\min}}{2}\right)\sin^w\left(\frac{\theta_{\min}}{2}\right).
\end{align}
We can evaluate the trace-norm distance with this inner product.


\end{document}